\documentclass[twocolumn,showpacs,groupedaddress]{revtex4}
\usepackage{graphicx}
\usepackage{epsfig}
\usepackage{subfigure}
\usepackage{amsmath}

\begin{document}

\newcommand{\ket}[1]{|#1\rangle}
\newcommand{\bra}[1]{\langle#1|}
\newcommand{\Tr}{\text{Tr}}

\title{
New Understandings of Quantum Mechanics Based on Interaction
}
\author{Tian-Hai Zeng}
\email{zengtianhai@bit.edu.cn}
\affiliation{Department of Physics, Beijing Institute of Technology, Beijing 100081, China}

\begin{abstract}
The interaction between two parts in a compound quantum system may be reconsidered more completely than before and some new understandings and conclusions different from current quantum mechanics are obtained, including a strict conservation law in the evolution in an isolated quantum system, new understandings of duality of particle and wave, measurement, and the principle of superposition of states, three laws corresponding to Newton's laws, new understanding of the uncertainty relation, support of the locality of Einstein \textit{et al.} and arguments against the non-locality of any entangled state, and a simple criterion of coherence which is obtained for experimenters to examine the correctness of the non-locality. These may make quantum mechanics be a bit more easily understood intuitively.
\end{abstract}

 \maketitle

\section{Introduction}\label{Introduction}

From the birth (1925-1926) of quantum mechanics to now, it has already produced some strange, mysterious or anti-intuitive superposed states of quantum systems, for examples, a pure state may be superposed by the ground state $|g\rangle$ and an exited state $|e\rangle$ of an isolated atom without interaction or interchanging energy with its outside, and a pure entangled state, which is also a superposed state and cannot be represented as a product of two wave functions describing two subsystems, may still maintain the entanglement after the interaction between the two parts ceases. The property of entanglement is called non-locality and considered as spooky action \cite{Tittel}.

In some cases, after the interaction between two parts of a compound system ceases, a subsystem is considered in a pure superposed state and disentangles with other subsystem. For example, an electron through double-slit is considered in a pure superposed state in standard textbooks of quantum mechanics; on the other hand, when an electron is going through double-slit, the interaction between the electron and the matter of the double-slit certainly exists, and the state of the two parts evolves into an entangled state, whether the interaction ceases or not, the entanglement should be maintained according to current quantum mechanics, then the electron itself is \textbf{not} in a pure superposed state! Therefore an absurd conclusion is obtained from quantum mechanics that a physical process may be considered to obtain two different results.

Until now, almost all authors of the books of quantum mechanics, for example, von Neumann \cite{Neumann}, Dirac \cite{Dirac} and Landau \textit{et al.} \cite{Landau}, thought that non-degenerate energy eigenstates, for example, $|g\rangle$ and $|e\rangle$ of an isolated atom without interaction with its outside, could be superposed, the reasons may be that not only a particle has wave superposition property according to de Broglie's assumption \cite{de Broglie} about matter wave, but also the wave function standing for the superposed state satisfies the Schr\"{o}dinger equation which is based on the matter wave property.

Although Feynman said \cite{Feynman} that no one can understand quantum mechanics, including the above strange states, and non-locality, many people have always their own understandings different from current points of quantum mechanics due to the points being not all satisfying. Those strange states and indigestible properties made Einstein \textit{et al.} \cite{Einstein} think that the theory of quantum mechanics is incomplete, and led to the famous argument of complete property of quantum mechanics between Einstein and Bohr \cite{Bohr}. The argument had been staying in philosophy until Bell gave an inequality or a theorem \cite{Bell}, which was based on hidden variable theory \cite{Bohm} and local reality theory \cite{Einstein}, trying to test the correctness of non-locality of entanglement in experiments and answer the issue of complete property. From then on, a large amount of investigations have been made to find evidence to prove the non-locality theoretically \cite{Clauser,Hardy,Greenberger} and experimentally \cite{Tittel,Aspect,Strekalov,Bouwmeester,Rowe}.

We do not know how profound the physical significance of principles of quantum mechanics is, but we may satisfy the understandings of it being a bit more profound than current. A free particle or an isolated quantum system is only an assumption, since it is tiny and always subject to the impact from background or heat-reservior. Therefore a quantum system always accompanies its outside, and there exists interaction between them. The interaction between two parts in a compound quantum system may be reconsidered more completely than before and some different understandings and conclusions from current quantum mechanics are obtained in this paper, including a strict conservation law in an isolated quantum system in the evolution (Sec.\ref{Conservation}), new understandings of duality of particle and wave (Sec.\ref{Duality}), measurement (Sec.\ref{Measurement}), and the principle of superposition of states (Sec.\ref{Superposition}), three laws corresponding to Newton's laws (Sec.\ref{Newton}), new understanding of the uncertainty relation (Sec.\ref{Uncertainty}), support of the locality of Einstein \textit{et al.} and arguments against the non-locality of any entangled state, and a simple criterion of coherence is obtained for experimenters to examine the correctness of the non-locality (Sec.\ref{Non-locality}). Section \ref{conclusion} is for the conclusions.

\section{Conservation Laws}\label{Conservation}

According to quantum mechanics, the conservation laws of energy, momentum and angular momentum hold only in the sense of a statistical average, not in the strict sense that an isolated quantum system (single particles or compound quantum systems) does not interchange these physical quantities with its outside in the evolution and maintains the conservation of the quantities at any time.

Perhaps most people prefer the conservation laws in the strict sense than in the sense of the statistical average, since the strict law does not contradict with the classical idea, that the interchanging of energy (momentum or angular momentum) is owing to interaction between two subsystems and then each of the quantities maintains conservation at any moment in the evolution; if a quantum system is isolated, i.e., there is no interchanging of the quantities with its outside, then the quantities will not vary. But we know that the principle of superposition of states and the uncertainty relation make one accept the conservation laws in an average sense. For example, an isolated atom, which is in ($|g\rangle$ + $|e\rangle$), maintains conservation of energy in the sense of a statistical average in evolution. My understandings of quantum mechanics, including Sec.\ref{Superposition}and Sec.\ref{Uncertainty}, may resume the conservation laws in a strict sense and there is no contradiction among them.

It is well known that the scattering of a photon and an electron, which compose an isolated compound system, obeys the conservation law of energy and matter and the conservation law of momentum all the time from the Compton scattering experiments \cite{Compton}. This may be explained as such that the interaction between the photon and the electron interchanges energy and matter, and momentum between them, and there is no interaction or interchanging these physical quantities with their outside. The strict conservation law of momentum and energy are often used in the process of quantum electrodynamics \cite{Greiner}, also due to the existence of an interaction between subsystems and interchanging the physical quantities in the process.

In quantum optics, a simple isolated compound system is composed of a single two-level atom and a single mode quantized field with an interaction between them, the wave function \cite{Walls} standing for the system state is

\begin{equation}\label{eq1}
  |\Psi_{af}\rangle=\frac 1 {\sqrt 2}(|g\rangle|n+1\rangle+|e\rangle|n\rangle),
\end{equation}
where $|n\rangle$ is an n-photon state. The photon energy equals the energy difference between the atomic exited and ground levels. The state Eq.(\ref{eq1}) is an entangled state and the wave function is a superposition of the two degenerate terms that their energies are equal, then it maintains the conservation of energy in the evolution of the isolated compound system all the time. The interaction plays a role of interchanging energy or other physical quantities between the atom and the field.

The conservation law may be comprehended as such that \emph{an isolated quantum system must maintain the conservations of energy, momentum and angular momentum at any time in the evolution, not in a sense of a statistical average.}

The penetration through a potential barrier to a particle seems to violate the conservation of energy. If we consider other matter which offers the potential barrier of interaction with the particle, the whole system evolves an entangled state and can keep the conservation of energy.

We think that those states of isolated systems, for examples, ($|g\rangle$ + $|e\rangle$) and ($|g\rangle|n\rangle$ + $|e\rangle|n+1\rangle)$, violate the strict conservation law in the evolution, then they do not exist in nature.

\section{New Understanding of Duality of Particle and Wave}\label{Duality}

The observations of some wave phenomena, for example, sound wave, water or liquid wave, elastic wave in solid, let me find out a common point that anyone of these waves has some interaction between particles. A wave is considered as some transmission of a vibration, which is viewed as a source of wave. Both transmission and vibration depend on respective interaction. It is easy to find out some interactions between particles in these waves, for example, the interaction between the molecules of atmosphere in sound wave. But the sound wave equation \cite{Feynman2},
\begin{equation}\label{eq2}
   \frac{\partial^2\xi}{\partial x^2}=\frac 1 {c^2}\frac{\partial^2\xi}{\partial t^2},
\end{equation}
may easily let one forget interaction. If there were no interaction, these wave phenomena could not appear in matter. So interaction is a requirement of producing or propagating these waves. Different interaction produces or propagates different wave. A complex wave, from a wave source, or produced by two or more waves meeting in some place, can be decomposed mathematically into some simple waves and viewed as a superposition of these simple waves. The complex wave must be corresponding to a superposition of some interactions.

For light, i.e., electromagnetic waves, when it meets double-slit or single-slit, the interaction between different parts of light or the interaction with the boundary matter of slit certainly exists, and then it behaves the property of wave, that is, the superposition of different parts. If there is no interaction, the pattern of the interference will disappear and it will not behave the property of wave. One of the most important light wave parameters, wavelength, cannot be measured without interaction, interference, for example, first used to measure the wavelength by Young in 1801 \cite{Born}. From the Maxwell equations:
\begin{equation}\label{eq3}
\begin{array}{l}
    \nabla \times \vec{E}=-\frac{\partial \vec{B}}{\partial t},\\
    \\
    \nabla \times \vec{H}=\frac{\partial \vec{D}}{\partial t},
\end{array}
\end{equation}
we may see that it is some interaction to vary the electric field ($\vec{E}$ and $\vec{D}$) or the magnetic field ($\vec{B}$ and $\vec{H}$), otherwise they are all static fields. In the wave equation of electromagnetic waves obtained from the Maxwell equations, it hides some interaction which makes the electric field and magnetic field vary. The viewpoint of electric and magnetic fields of Faraday is that they are all matters. The change of fields in continuous electromagnetic waves may be some action on charges, for example, antenna of radiation, and some interaction among the matters. If a radiation is a pulse, then a photon is produced. We consider one photon passing through double-slit or single-slit, the explanation of its probability wave superposition property is also due to the interaction between single photons and the boundary matter of slit. \emph{So the wave property of light comes from its particle property plus some interaction, and particle property is more fundamental than wave property. A single photon itself has no probability wave property and the principle of superposition of its probability waves could not hold without interaction.}

For a material particle, an electron, for example, the typical experiments of proving its wave property are the crystal diffraction \cite{Davisson} and the double-slit interference \cite{Zhang}. The pattern of the diffraction or the interference, the characteristic of the wave superposition property, is also due to the existence of the interactions between an electron and the crystal or the double-slit in the two experiments. If the interaction ceases, or the electrons are far from the crystal or the double-slit, the pattern will disappear. \emph{An electron may interact with its electromagnetic field, then its wave property may be intrinsic. If there is no interaction, a single neutral material particle itself has no wave property and the principle of superposition of its probability waves could not hold either.}

A material particle is subject to tiny action and there exists no area where any interaction does not exist, the vacuum, for example, can not be obtained, then the background field always interacts with the material particle considered, therefore its wave property may be considered intrinsic too. So we can understand that the greater the energy of a particle, the smaller the impact to it produced by background field or photon, then the shorter the de Broglie wavelength of it. Although the background field is difficult to be eliminated, the interaction between two particles can be controlled to become zero, then the superposition of the waves about the system of the two particles will not exist in nature.

\section{New Understanding of Measurement}\label{Measurement}
The mainstream point of measurement is such as pointed out by Dirac \cite{Dirac} that, \emph{\textquotedblleft From physical continuity, if we make a second measurement of the same dynamical variable $\xi$ immediately after the first, the result of the second measurement must be the same as that of the first. Hence after the first measurement has been made, the system is in an eigenstate of the dynamical variable, the eigenvalue it belongs to being equal to the result of the first measurement\textquotedblright.} This is different from the point of Landau \textit{et al.} \cite{Landau} that, \emph{\textquotedblleft after the measurement, however, the electron is in a state different from its initial one, and in this state the quantity \emph{f} does not in general take any definite value. Hence, on carrying out a second measurement on the electron immediately after the first, we should obtain for \emph{f} a value which did not agree with that obtained from the first measurement\textquotedblright.}

Since Landau \textit{et al.} considered that the measurement of a microscopic system needs some interaction between an apparatus and the measured system, and the compound system evolves an entangled state, I think that their point is a bit better than that of Dirac or the mainstream point of measurement. The common of the above two points is that the measured values are all eigenvalues. But I think that most eigenvalues are not observables, which is explained below.

If not measuring a microscopic system or no change in the apparatus, we even cannot know whether the system exists. Therefore a successful measurement of a microscopic system may be read from the change of the apparatus and the system must be also changed. This may be a fundamental of measurement of microscopic world and is very different from that of macroscopic world, which is a comparison with a standard apparatus without disturbing the system.

Up to now, only the difference of energy eigenvalues could be measured. The energy of a photon (or other quantized field) can be measured if the photon is absorbed or destroyed, for example, atomic spectrum. In addition, when we mention the potential energy of a particle in a field, only its difference has physical meaning. An eigenvalue of a material particle in an energy eigenstate may be measured as such that the final state should have zero energy, then the corresponding difference between the energy eigenvalue and zero could be read from the change of the apparatus. However, the final state may not be certainly in the state of zero energy, then the eigenvalue may not be certainly measured.

The above observations let me think that the measured physical quantities of a system may be divided into three kinds that we call them: \emph{inherent vector} (for example, spin and photon polarization, the magnitude of it is constant along any direction, and its direction or eigenvalue can be directly measured), \emph{non-inherent vector} (for example, velocity, momentum and angular momentum, the magnitude and direction of it are alterable) and \emph{scalar quantity} (for example, coordinate, potential and kinetic energy).

A measurement may be completed in an interval of time $\Delta t$ or of space $\Delta x$, and alter the state of the system considered. One direction of an inherent vector can be considered as an eigenvalue of corresponding eigenvector or eigenstate, and can be directly obtained in measurement, and left it being in the direction or eigenvector, while the whole state of the particle or other quantity must be changed. For example, if an electron is initially in one spin eigenstate and is measured in the Stern and Gerlach experiment, the eigenvalue or direction of spin is measured and the electron still stays in the eigenstate, whereas the direction of the electron motion is changed. For a measured value of other quantities, energy and momentum, for example, only the difference of two eigenvalues can be obtained, not one of their eigenvalues, therefore we can say that \emph{these eigenvalues themselves, other than inherent vector, are not observables, and the system is not in the initial eigenstate after the result is read.}

The statistical explanation of wave function of an isolated compound system then is slightly changed as such that one of the subsystems is only in an eigenstate with some probability, and only the direction of an inherent vector or eigenvalue can be measured, while the other eigenvalues cannot. This is different from that in quantum mechanics, that all eigenvalus can be obtained with some probabilities. This is also different from the idea of the physical reality \cite{Einstein} that, \emph{\textquotedblleft if, without in any way disturbing a system, we can predict with certainty the value of a physical quantity, then there exists an element of physical reality corresponding to this physical quantity\textquotedblright.}

Some change of a compound system, which is brought by an apparatus in a process of measurement, also brings a break or disappearance of the interaction between parts. It is the break or disappearance of the interaction in a compound system that brings the collapse of the wave function (reduction of wave packet) or disentanglement, and evolves a new entangled state if new interaction appears.

\section{New Understanding of the Principle of Superposition of States}\label{Superposition}

In his book, Zhang \cite{Zhang} points out that interactions always have the effect of non-linearity, which conflicts to the linearity of the principle of superposition of states; the interaction potential in the Schr\"{o}dinger equation has been treated with \emph{external field approximate} and then has an approximate linearity. This approximate linearity may suit for the linearity of the principle. Therefore the exact consideration of non-linearity of interaction must destroy the linearity of the principle, and the linearity may make some superposed states deviate real states and become strange. But a large number of results obtained from quantum mechanics with the linearity accord with results of experiments. So this approximate linearity is good enough, and the non-linearity of interactions cannot be used to explain the strange superposed states in Sec.\ref{Introduction}.

The hidden variable theory has been produced to explain the strange superposed states, but hidden variable is still mysterious up to now. The exhaustive explanations of the strange superposed states may be very difficult, but we shall satisfy the decrease of the mysterious extent of the strange states with the following understanding.

The Schr\"{o}dinger equation of an isolated compound system of two particles is
\begin{equation}\label{eq4}
    i\hbar\frac{\partial\Psi}{\partial t}=(\frac{\hat{p}_1^2}{2m_1}+V(r_{12})+\frac{\hat{p}_2^2}{2m_2})\Psi,
\end{equation}
where $\Psi$ is the total wave function of the two particles, $\frac{\hat{p}_1^2}{2m_1}$, $\frac{\hat{p}_2^2}{2m_2}$ and $V(r_{12})$ are the kinetic energy operators of the particles and the interaction potential energy between them, respectively. We suppose that $\{\psi_{1i}\}$ and $\{\psi_{2j}\}$ are the complete collections of the kinetic energy eigenstates of the particles, and $\{K_{1i}\}$ and $\{K_{2j}\}$ are the kinetic energy eigenvalues, respectively. According to quantum mechanics, $\Psi$ can be expanded by the collection $\{\psi_{1i}\psi_{2j}\}$ as
\begin{equation}\label{eq5}
    \Psi=\sum_{ij}\alpha _{ij}\psi_{1i} \psi_{2j}.
\end{equation}
If $\Psi$ has two or more terms, it is an entangled state. Different term in  $\Psi$ may have different kinetic energy $(K_{1i}+K_{2j})$, while the total energy, $E=K_{1i}+K_{2j}+V_{1i2j}$ corresponding to the different term $\psi_{1i} \psi_{2j}$ may maintain conservation in the evolution, due to the interaction interchanging energy within these three parts of the isolated system. Those states, in which total energies of different terms are different, should not exist in nature.

If $V(r_{12})=0$, $\Psi$ still satisfies the Schr\"{o}dinger equation (\ref{eq4}), and maintains entanglement according to quantum mechanics. This is a strange state that it may be in different total energy eigenstates and each one of the two isolated particles may be in different kinetic energy eigenstates without interaction with its outside, which contradicts the strict conservation law of energy. According to our observation, if there is no interaction or interchanging energy with each other, the principle of superposition of waves will not hold, then there is only one term in $\Psi$, i.e., there is no entanglement. If we do not know the classical information of preparation of the two particles, we have to describe the state in a mixed state
\begin{equation}\label{eq6}
    \rho=\sum_{ij}p_{ij}|\psi_{1i} \psi_{2j}\rangle\langle \psi_{1i} \psi_{2j}|.
\end{equation}
It is a disentangled mixed state. The wave function $\Psi$ can also be expanded by the complete collection of other eigenstates, momentum of one dimension, for example,
\begin{equation}\label{eq7}
    \Psi=\sum_{ij}\beta_{ij}\psi^p_{1i} \psi^p_{2j}.
\end{equation}
The momentum may be interchanged between the two particles, so the total momentum $p_{1i}+p_{2j}$ corresponding to the different term $\psi^p_{1i} \psi^p_{2j}$ should be conserved, or these terms are degenerate for momentum.

The Schr\"{o}dinger equation,
\begin{equation}\label{eq8}
    i\hbar\frac{\partial\psi}{\partial t}=(-\frac{{\hbar}^2}{2m}\Delta+V)\psi,
\end{equation}
has been used to describe the dynamic evolution of a particle as a probability wave by a wave function $\psi$, and generalized to micro-systems. The potential $V$ is obviously expressed in the equation, and it is considered as an external field approximation \cite{Zhang}, and a part interacted with the particle or the micro-system is neglected in Eq.(\ref{eq8}). If there is no $V$, a particle or a micro-system should be isolated and there is no energy, momentum and angular momentum exchanging with its outside. \emph{So an isolated particle or micro-system should have definite energy, momentum and angular momentum, and no state of a particle or a micro-system is any superposition of different eigenstates of the three physical quantities.}

On the electron double-slit experiment discussed by wave function, we suppose that $|\psi_1\rangle$ denotes the state of the electron passing through the slit 1 and $|M_1\rangle$ denotes the state of the slit matter corresponding to $|\psi_1\rangle$, $|\psi_2\rangle$ and $|M_2\rangle$ are similar. Although the interaction potential between the slit matter and the electron passing through the slits may be very complicated, and has not been expressed in the Hamiltonian, it always exists. We may explain it as such that the compound system is in some state
\begin{equation}\label{eq9}
   |\Psi_{es}\rangle=\alpha|\psi_1\rangle|M_1\rangle+\beta|\psi_2\rangle|M_2\rangle,
\end{equation}
which is an entangled state, and the electron will be in an approximately superposed state, i.e., the popular one in quantum mechanics
\begin{equation}\label{eq10}
   |\Psi_{e}\rangle=\alpha|\psi_1\rangle+\beta|\psi_2\rangle,
\end{equation}
if the states $|M_1\rangle$ and $|M_2\rangle$ are considered same approximately, therefore we can also use the state (\ref{eq9}) to explain the pattern of interference of the electron through double-slit as well as the state (\ref{eq10}). If the state (\ref{eq9}) is not considered as approximately equal to the state (\ref{eq10}), the pattern will be different. This explanation is a bit different from that in current quantum mechanics. The state (\ref{eq9}) is less mysterious than the state (\ref{eq10}), we shall satisfy the decrease of the mysterious extent.

In quantum optics, the wave function (\ref{eq1}) is a superposition of the two terms that their energies are equal and then it maintains the strict conservation of energy. The atom will be in an approximately superposed state
\begin{equation}\label{eq11}
  |\Psi_{a}\rangle=\frac 1{\sqrt 2}(|g\rangle+|e\rangle),
\end{equation}
if $n$ is large. Otherwise it may be in a mixed state, which is described by a reducible density operator by tracing out the part of photons, and not in a pure superposed state of its two energy eigenstates. If the state (\ref{eq11}) is exact, a novel cat state, similar as (\ref{eq11}), must be stranger than the Schr\"{o}dinger's cat state, similar as (\ref{eq1}). The novel cat state is in a directly superposed state of dead cat and live cat states, or a cat is in Wheeler home and in Einstein home at same time \cite{Aczel} and there is no reason, while if the Schr\"{o}dinger's cat is live, the reason is due to the cover of the toxicant bottle unopened. Therefore the state (\ref{eq10}) or (\ref{eq11}) is only an approximately superposed state and still relate with interaction, that is, the exact superposed state (\ref{eq10}) or (\ref{eq11}) is only an assumption and will not exist in nature. If the interaction ceases, the state (\ref{eq9}) or (\ref{eq1}) will be the state $|\psi_1\rangle|M_1\rangle$ (or $|\psi_2\rangle|M_2\rangle$) or $|g\rangle|n+1\rangle$ (or $|e\rangle|n\rangle$) and the two parts will not be in an entangled state. This agrees with the locality viewpoint by Einstein \textit{et al.} \cite{Einstein} that, \emph{\textquotedblleft at the time of measurement the two systems no longer interact, no real change can take place in the second system in consequence of anything that may be done to the first system\textquotedblright.}

The other two similar examples are coherent and squeezed states \cite{Walls} of radiation fields. They have been expressed respectively in different superposition of photon number states with different probability amplitudes, or each one is in a pure superposed state. These states cannot maintain the strict conservation of energy in the evolution. When these states are preparing, the field and the apparatus have interaction and evolve an entangled state. After the interaction ceases, the entanglement can be maintained according to quantum mechanics, then the field cannot be in a pure superposed state of photon number states, i.e., coherent or squeezed state cannot exist in nature. This prepared field may be some photon number state and can be described as a mixed state. If the interaction does not cease and the states of the apparatus are approximately considered same, then we can obtained approximate coherent or squeezed state. The vacuum fluctuation may also be explained as a result of the states of other matter being approximate same.

There are a large amount of approximate solutions of the Schr\"{o}dinger equation of systems in quantum mechanics since exact solutions are difficult to be obtained, while the states (\ref{eq10}) and (\ref{eq11}) are viewed as exact solutions, so why their physical meaning are difficult to be understood is because of neglecting the state of other matter (or external field approximate) and even the interaction between the matter and the system considered.
However, it was probably this neglect that brought the hidden variable presented for explaining some strange properties of a quantum system and even all other matter is considered if the decoherence of a superposed state is discussed, the two extremities make quantum mechanics more indigestible.

The above cases hint us that the understanding of the principle of superposition of states in quantum mechanics may be changed as such that, \emph{in non-relativistic quantum mechanics, only a superposed state of inherent vector (spin and photon polarization) of a single particle could exists in physics, other superposed states exist only in compound systems with interaction between subsystems and are entangled states, interaction and strict conservation law are new constrain conditions.}

A state has been believed to be mathematically expanded as a superposed state of eigenstates of a conserved physical quantity, but it has not been exactly proved. We think that the expansion may not be a single, i.e., a state may be mathematically expanded as a superposed state of not only non-degenerate eigenstates of a conserved physical quantity, but also degenerate eigenstates, for example, two terms in Eq.(\ref{eq1}), the latter has physical meaning while the former no. The degenerate states of a system may be those of energy, or momentum, or angular momentum.

The Schr\"{o}dinger equation of a hydrogen atom system is transformed to a single particle's equation in an equivalent potential field. We can solve the equation to obtain the energy eigenstates and eigenvalues of the atom. It can be in some energy eigenstate, but it cannot be in a superposed state of non-degenerate energy eigenstates according to our understanding. When it is in an energy eigenstate, the hydrogen atom system composed of an electron and a proton can be in an entangled state, therefore they may be in different states due to the interaction interchanging energy among three parts (the kinetic energies of the electron and the proton, and interaction potential energy). The energy of the atom can also be divided into two parts: one is the sum of the kinetic energy of the electron and interaction potential energy, and the other is the kinetic energy of the proton. Momentum or angular momentum can also be interchanged between the electron and the proton.

\section{Three Laws Corresponding to Newton's Laws}\label{Newton}

Since the operators of energy, momentum and angular momentum of a free or isolated particle commute with each other, they have common eigenfunctions. By use of our understanding, we may obtain a law corresponding to the Newton's first law that \emph{a free particle must be in an eigenstate (an extrapolated wave function, that is, a plane wave) having definite energy, momentum and angular momentum in some inertial reference frame.} But we do not know what state the free particle is in after it is prepared, so we have to describe it in a mixed state of other type (described by more than one wave function or also by a density operator) and not in a superposed state.

\emph{The Schr\"{o}dinger equation may be corresponding to the Newton's second law,} for they are all dynamic equations of their respective system.

In a compound system with interaction, the states of all parts should be considered in the Schr\"{o}dinger equation for exactness or understanding. \emph{The principle of superposition of states in a compound system with new understanding may be corresponding to the Newton's third law}.

\section{New Understanding of the Uncertainty Relation}\label{Uncertainty}

The uncertainty relation was obtained due to the fact that the non-commuting operators, momentum $p_x$ and coordinate $x$, for example, corresponding to two physical quantities have no common eigenfunction. If the system is in an eigenstate of one operator, it cannot be in an eigenstate of the other. This led Einstein et al. present the physical reality viewpoint above. But we can understand it as such that the momentum $p_x$ of a free particle keeps definite value in any coordinate $x$, i.e., when $p_x$ is an eigenvalue, the particle has no a definite coordinate; therefore it is not strange. But self-contradiction appears below.

According to the mainstream point of measurement pointed out by Dirac \cite{Dirac}, a measurement value of a system is one of the eigenvalues, with some probability, associated with the eigenfunctions in the wave function, and the state left is the eigenstate; if the quantity is measured immediately, the energy, for example, the same eigenvalue may be obtained, the difference value of two results $\Delta E=0$, but the interval $\Delta t$ between two measurements, according to the point of Landau \textit{et al.} \cite{Landau}, is not infinite, then
$\Delta E\cdot\Delta t=0$. Therefore the measurement idea of quantum mechanics contradicts with the uncertainty relation of energy-time $\Delta E\cdot\Delta t\geq \hbar/2$.

If we associate the uncertainty relation with our new understanding of measurement above, the difference of energy or momentum is measured at least one photon's. Then the uncertainty relation of energy-time can be re-explained as the following. If the energy of one photon emitting from an atom is measured, the atom has decreased the same energy. Therefore,
\begin{equation}\label{eq12}
  \Delta E'=h \nu =\frac h T,
\end{equation}
Where, $h$, $\nu$ and $T$ are Planck's constant, frequency and \textit{period} of the photon. The time needed in one measurement may be equal or great than the period, i.e., $\Delta t'\geq T$, then we obtain $\Delta E'\cdot\Delta t'\geq h$, this is different from the meaning and the formula of the uncertainty relation, $\Delta E\cdot\Delta t\geq \hbar/2$, in which $\Delta E$ represents the difference between two measured eigenvalues and $\Delta t$ a time interval of two measurements; \emph{while the $\Delta E'$ and $\Delta t'$ come from only one measurement, not two.} Similarly, in $x$ direction the least change of momentum of a particle is also a photon's
\begin{equation}\label{eq13}
  \Delta p_x'=h/cT
\end{equation}
where $c$ is the light speed. A measurement may be completed in the extent $\Delta x'\geq cT$, then $\Delta x'\cdot\Delta p_x' \geq h$, which is also different from the meaning and the formula of the uncertainty relation, $\Delta x\cdot\Delta p_x \geq \hbar/2$.

\section{Support of the locality and Arguments against Non-locality of Entangled State}\label{Non-locality}

According to our understandings of wave and the principle of superposition of states, the entanglement in a compound system is produced and maintained by interaction among the parts. The principle of indistinguishability of identical particles, which is based on exchange interaction, seems to be other reason to produce and maintain entanglement without interaction potential energy in the Hamiltonian of the system. But a pure entangled state and its corresponding mixed state (for example, Eqs.(\ref{eq14}) and (\ref{eq16}) below) are all fit for the principle, since the expressions of the two states are completely equivalent respectively by exchanging two particles, and then equivalent in physics, so we think that exchange interaction is imaginary and different from other real interactions in Hamiltonian, and then not the reason of producing and maintaining entanglement. If the interaction between two identical particles, other than the exchange interaction, ceases, the superposed state of a compound system will collapsed, i.e., entangled state will not exist.

Having reinvestigated Bell's theorem (inequality) \cite{Bell} and later ones \cite{Clauser,Hardy,Greenberger} and some related experiments \cite{Tittel,Aspect,Strekalov,Bouwmeester,Rowe}, we cannot find out that any given pairs of particles without entanglement was used in the experiment same as the same particles with initial entanglement, then there is no comparison of measurement results of the two states, furthermore no average result of coherent probability surpasses 75\% (the ideal classical coherent probability explained below).

In deduction of his inequality starting from hidden variable theory and local reality theory, Bell had used the formula of coherence of an entangled spin state of two electrons, $A(\vec{a},\lambda)=-B(\vec{a},\lambda)$, which is a result of quantum mechanics and cannot be obtained from the formulae $A(\vec{a},\lambda)=\pm 1$ and $B(\vec{b},\lambda)=\pm 1$, in the case there is no interaction between the two electrons. If the formula $A(\vec{a},\lambda)=-B(\vec{a},\lambda)$ comes from some experiment, then the non-locality of an entangled state has been proved and we do not need the inequality.

In all experiments to test Bell's theorem and later ones, two loopholes \cite{Rowe}, that one is low detection loophole and the other is locality or lightcone loophole, which is about two parts of an entangled state being no spacelike separate associated with measurement, cannot be closed at same time. Other type experiment for testing the non-locality of entangled state is quantum ghost interference \cite{Strekalov}. In the experiment, very few e-ray photons pass through the double slits to photon counting detector, meanwhile many o-ray photons reach the other detector, then the output pulses of the detectors, sent to a coincidence circuit with $1.8$ nsec coincidence time window, may not be a pair of initially entangled photons. So this experiment is still not enough to prove the non-locality of an entangled state.

In the following, we compare the calculations of coherent probability of an entangled state, which is assumed to maintain the entanglement when the interaction between two parts ceases, and that of some probable disentangled mixed states. We first consider two distant identical particles in different energy eigenstates $|0\rangle$ and $|1\rangle$, respectively. Suppose that the interaction between them is only in a very small area, and when they approach and interact with each other, their state will evolve in an entangled state
\begin{equation}\label{eq14}
 |\Psi_{12}\rangle=(|10\rangle+|01\rangle)/\sqrt 2 .
\end{equation}
According to quantum mechanics, the state will maintain the entanglement when they apart from each other and the interaction between them ceases, and the state form can be rewritten as
\begin{equation}\label{eq15}
  |\Psi_{12}\rangle=(|AA\rangle-|SS\rangle)/\sqrt 2
\end{equation}
where $|A\rangle=|0\rangle+|1\rangle)/\sqrt 2$ and $|S\rangle=|0\rangle-|1\rangle)/\sqrt 2$. In the experiment \cite{Rowe}, the states of one particle superposed by energy eigenstates, similar as $|A\rangle$ and $|S\rangle$, are considered to be produced by Raman beam. But according to our understanding, the energy state of the single particle and Raman beam is an entangled state and the single particle itself is not a pure superposed state, then the entangled state in Eq.(\ref{eq14}) cannot be written as Eq.(\ref{eq15}). When the interaction of the two identical particles ceases, the entangled state collapses into the state $|10\rangle$ or $|01\rangle$, which can be described by density operator form
\begin{equation}\label{eq16}
  \rho=\frac 1 2 (|10\rangle\langle 10|+|01\rangle\langle 01|),
\end{equation}
 a disentangled mixed state. We cannot distinguish the mixed state and the entangled state by measuring difference of energy of two eigenstates of a single particle or even energy eigenvalues with the measurement concept of quantum mechanics, that is, if we measure their energies, we may all obtain 100\% coherent probability, therefore the non-locality of entangled state of this type cannot be proved experimentally.

Cohen \cite{Cohen} discovered that a mixed state of two same subsystems could be written as
\begin{equation}\label{eq17}
  \rho_1=\frac 1 2 (|00\rangle\langle 00|+|11\rangle\langle 11|),
\end{equation}
or
\begin{equation}\label{eq18}
  \rho_2=\frac 1 4 (|00\rangle+|11\rangle)(\langle 00|+\langle 11|)+\frac 1 4 (|00\rangle-|11\rangle)(\langle 00|-\langle 11|),
\end{equation}
due to $\rho_1=\rho_2$ in mathematically. Cohen thought that there exists hidden entanglement. But we think that the mixed state will be expressed by Eq.(\ref{eq17}) (no entanglement) if the interaction between the two particles ceases after the system state is prepared. If some interaction between them exists and $|0\rangle$ and $|1\rangle$ represent the degenerate energy eigenstates of a subsystem, the mixed state may be expressed by Eq.(\ref{eq18}), which exists entanglement. So this is also a defect of density operator that it expresses two different mixed states, entangled and disentangled ones.

Next, we consider the polarized (inherent vector) state of two photons. If a pair of photons, with the horizontal state $|H\rangle$ and vertical state $|V\rangle$ respectively, enter into a beam splitter and interact, the two photons may be in the singlet state
\begin{equation}\label{eq19}
 (|HV\rangle-|VH\rangle)/\sqrt 2= (|+-\rangle-|-+\rangle)/\sqrt 2,
\end{equation}
where $|+\rangle=\cos\gamma|H\rangle+\sin\gamma|V\rangle$ and $|-\rangle=-\sin\gamma|H\rangle+\cos\gamma|V\rangle$ represent $\gamma$ and $\gamma+90^\circ$ polarized photons with the angle $\gamma (0\sim 90^\circ)$ between $|H\rangle$ and $|+\rangle$. After they come out of the beam splitter, the interaction disappears and the entangled state collapses in a mixed state according to our understanding, but their state maintains entanglement in Eq.(\ref{eq19}) according to current quantum mechanics. We do not know the scheme of collapse of wave function, we guess that the first probable mixed state may be in an ensemble of $|+-\rangle$ or $|-+\rangle$ in different angle $\gamma$ with identical probability density, and the second may only be in one of the states $|HV\rangle$ or $|VH\rangle$ with same probability. If we select two measurement bases $\{|H\rangle,|V\rangle\}$ and $\{|+\rangle,|-\rangle\;\;\text{in any angle}\;\;\gamma\neq 0\}$, all results will be 100\% coherent probability for entangled state, that the polarizations of the two photons must be perpendicular. But, if the measured state is a disentangled mixed state, our calculation in an average is 75\%, for the first mixed state by the same measurement way as above, that is, if one photon is measured in state $|+\rangle$, the other photon is measured in state $|-\rangle$ with 75\%, and 100\%$\sim$50\% with $\gamma=0\sim 45^\circ$ for the second case, and the average coherent probability is also 75\%. So we can distinguish experimentally which state, entangled or disentangled, the measured coherent probability of initially entangled state belongs to if the probe efficiency is high enough.

We may prepare some pairs of particles without entanglement and do same experiments as measuring the initially entangled states of same particles, then we can compare the measurement results to discover whether the initially entangled state has been disentangled. If the ratio of results of coherent probability is approximately 3:4, the latter states maintain their entanglement; if the ratio of that is near equal, then the latter states have been disentangled. This way may be used to test the non-locality of an entangled state in the case of low probe efficiency.
\vspace {0.1cm}

\section{conclusions}\label{conclusion}

In this paper, I consider the interaction in a quantum system more completely than before, and produce some new understandings and conclusions of quantum mechanics. These may make quantum mechanics be a bit more easily understood intuitively and some strange properties will not appear, for example, a superposed state of a free particle, except inherent vector, and the non-locality of an entangled state will not appear. The new understandings and conclusions are:

An isolated quantum system must maintain the conservation of energy, momentum and angular momentum at any time in the evolution, not in a sense of a statistical average.

If there is no interaction, wave will not appear and the principle of superposition of waves could not hold.

The measured value must be successfully read from the change of the apparatus state and this must change the state of the system.

In non-relativistic quantum mechanics, only a superposed state of inherent vector (spin and photon polarization) of a single particle could exists in physics, other superposed states exist only in compound systems with interaction between subsystems and are entangled states, interaction and conservation law are new constrain conditions.

There are three laws in quantum mechanics similar with Newton's laws.

The uncertainty relation results from only one measurement, not two.

The coherent probability of an entangled state is 100\%, and greater than that of a disentangled mixed state, in average 75\%.

Therefore the complete consideration of interaction may make the understandings of quantum mechanics a bit more profound than before, and also produce a viewpoint same as the locality. We may not need for the moment some hidden variable theory and a complete theory of quantum mechanics that Einstein believed \cite{Einstein} to be produced in future.

\vspace{-0.2cm}
\begin{acknowledgments}
\vspace{-0.2cm}
I acknowledge my colleagues: Jian Zou, Feng Wang, Bin Shao, Xiu-San Xing and Jun-Gang Li for discussions or comments, thank my former classmates: Li-Fan Ying and Gui-Qin Li for useful helps, especially thank my former teachers: Pei-Zhu Ding and Shou-Fu Pan for their encouragements.
\end{acknowledgments}

\end{document}